\documentclass[pra,aps,onecolumn,notitlepage, superscriptaddress]{revtex4-1}
\usepackage[T1]{fontenc}

\usepackage{amsmath,amssymb}
\usepackage{amsfonts}
\usepackage{epsfig,pstricks,graphicx}
\usepackage{bm}
\usepackage{color}
\newcommand{\F}{\mathcal F}

\newcommand{\FQ}{\mathcal F^Q}
\newcommand{\U}{\mathcal{U}}
\newcommand{\ket}[1]{\left | #1 \right\rangle}
\newcommand{\marcin}[1]{{\color{black} #1}}

\newcommand{\marcinNew}[1]{{\color{black} #1}}

\usepackage{xcolor}
\usepackage[normalem]{ulem}
\begin{document}

\title{Simultaneous estimation of multiple phases in generalised Mach-Zehnder interferometer}

\author{Marcin Markiewicz}
\email{marcin.markiewicz@ug.edu.pl}
\affiliation{International Centre for Theory of Quantum Technologies, University of Gdańsk, 80-308 Gdańsk, Poland}

\author{Mahasweta Pandit}
\affiliation{Institute of Theoretical Physics and Astrophysics, Faculty of Mathematics, Physics and Informatics,
University of Gdańsk, 80-308 Gdańsk, Poland}

\author{Wiesław Laskowski}
\affiliation{International Centre for Theory of Quantum Technologies, University of Gdańsk, 80-308 Gdańsk, Poland}
\affiliation{Institute of Theoretical Physics and Astrophysics, Faculty of Mathematics, Physics and Informatics,
University of Gdańsk, 80-308 Gdańsk, Poland}

\begin{abstract}
In this work we investigate the problem of simultaneous estimation of phases using generalised three- and four-mode Mach-Zehnder interferometer. In our setup, we assume that the phases are placed in each of the modes in the interferometer, which introduces correlations between estimators of the phases. These correlations prevent simultaneous estimation of all these phases, however we show that  we can still obtain the Heisenberg-like scaling of precision of joint estimation of any subset of $d-1$ phases, $d$ being the number of modes, within completely fixed experimental setup, namely with the same initial state and set of measurements. Our estimation scheme can be applied to the task of  quantum-enhanced sensing in  three-dimensional interferometric configurations.
\end{abstract}

\maketitle

\section*{Introduction}

Multiparameter estimation is a rapidly growing field of quantum metrology \cite{Yue14, SzczykulskaRev, Ragy16, Gessner18, Liu19, Gorecki20, GeometricRev, PhotonicRev}, in which an initial quantum state undergoes an evolution which depends on several parameters, and the final task of the process relies on estimating these parameters based on measurement statistics of the final state with as low variance as possible. Multiparameter estimation encounters specific difficulties, which do not occur in the single-parameter case \cite{Toth14, Demkowicz15}. One of the most significant difficulties is the possibility of an occurrence of correlations between the estimators corresponding to different parameters, which could decrease the overall precision of joint estimation \cite{Ragy16}. Also, in this regard, it is worth mentioning that in \cite{Goldberg} it has been shown that the number of simultaneously estimatable parameters reduces when an external reference mode is absent. The main tool used to evaluate the precision of multiparameter estimation is the generalisation of the Quantum Fisher Information (QFI) \cite{Helstrom76} to the multiparameter case, known as the Quantum Fisher Information Matrix (QFIM) \cite{Liu19}. For sufficiently uncorrelated parameters the QFIM is invertible and the covariance matrix of the estimated parameters is bounded from below by the inverse of the QFIM. This bound is a multiparameter version of the Quantum Cramer-Rao bound (QCRB) \cite{Ragy16}. The elements of the QFIM depend on the size of the initial probe state. If the state is $N$-partite the elements of QFIM can depend at most quadratically on $N$, which is known as the Heisenberg-like (HL) scaling \cite{HL, HLUniv}.
Several works were showing that the Heisenberg-like scaling of precision of estimation of all the parameters is possible for an entangled input state and some measurement strategy, which in principle demands the use of arbitrary multiports \cite{Humphreys13, Yue14, Knott16, Liu16, Zhang16, Gessner18}.

\marcin{
In this work we state the problem of simultaneous estimation of multiple phases using $3$- and $4$-port generalised Mach-Zehnder interferometer \cite{Brougham2011}. We consider a scenario in which the phases are placed within each of the internal ports of the $d$-mode interferometer (see Figure 1a), and the task is to simultaneously estimate any $(d-1)$-element subset of them, whereas the remaining one is known, and serves as a phase reference.

Note that such configuration implies that the phases are strongly correlated, and the QFIM for all the phases in the interferometer is singular. The singularity of the QFIM reflects the impossibility of simultaneous estimation of all the phases without an \emph{external reference mode} \cite{Jarzyna12,Goldberg, Ataman20}.

We show that with the use of a fixed initial entangled probe state and a fixed interferometer one can obtain the Heisenberg-like scaling of precision of simultaneous estimation of any $(d-1)$-element subset of the phases, without any change in the setup. This means that we have both the same initial state as well as the same set of local measurements when estimating each subset.}

In a typical approach to quantum phase estimation one treats all the unitary part \emph{before} the phaseshifts as a preparation of the initial state, whereas all the part \emph{after} them is treated as an implementation of the measurement. In this work we apply a different point of view, treating the entire interferometer as a fixed single unitary operation, with the aim of investigating the metrological properties of a generalised Mach-Zehnder interferometer as a whole.

In our setup (see Figure 1) we use only fixed symmetric multiports and an $N$-partite entangled state which is equivalent to a GHZ state \cite{GHZ} of $d$-level systems via local unitary transformation specified by an additional symmetric multiport. To perform a detailed analysis of precision of estimation we develop an analytical description of a generalised 3- and 4-mode Mach-Zehnder (MZ) interferometer \cite{Brougham2011} composed of two symmetric multiports intertwined with single-mode phase shifters using the Heisenberg-Weyl operators. We found that the entire evolution of the quantum state in such interferometer is generated by generalised unitary Pauli $Y$ operators, which is analogous to the original two-mode case.
\marcin{This approach allows us to analytically assess the estimation precision of any $(d-1)$-element subset of phases by calculating the inverse of the classical Fisher Information Matrix\cite{Ragy16}. We show that even though we use the same initial state and the same measurement for estimation of all the subsets of phases, we are still able to obtain the Heisenberg-like scaling of precision of estimation of each of the subset. }

Our setup can be presented in two configurations which are equivalent from the estimation perspective. The first configuration, which we utilize for detailed analytical discussion of the estimation precision, consists of $N$ local MZ interferometers in a star-like configuration (see Figure 1b). Further, we present much more experimentally feasible single-interferometer version of the setup (just a single interferometer depicted in  Figure 1a), which uses the so-called NOON states \cite{NooNGen} as the initial probe states.

 There are several works discussing multiphase estimation with symmetric multiports \cite{Spagnolo12, Ciampini16, You17, Polino19, Li20}. They assume that the number of parameters to estimate is lower than the number of modes (in order to perform \emph{simultaneous estimation}) and also that the initial state \marcinNew{contains small definite number of photons}, which makes impossible the discussion of scaling of precision with the size of the input state. Our analysis differs from these works in the following points:\marcin{ (i) we assume the phases are placed in each mode of the interferometer, and demonstrate the possibility of obtaining the Heisenberg-like scaling of precision of simultaneous estimation of any $(d-1)$-element subset of the phases in a completely fixed interferometric setup,} (ii) we are not using the methods of QCRB, which in our case would not guarantee achievability of the Heisenberg-like scaling with the same fixed setup; instead we use directly the method of classical Fisher-information-based analysis on the basis of output probabilities, which do not raise the questions of achievability, as no implicit optimisation over measurement procedure is included, (iii) we assume the input state with arbitrary number of photons which allows directly for asymptotic scaling discussions.

As a side result, we show that contrary to previous statements \cite{Jex95} it is possible to construct a 5-mode fully symmetric multiport, the evolution of which is generated by a symmetric Hamiltonian.

\begin{figure}[ht]
    \centering
    \includegraphics{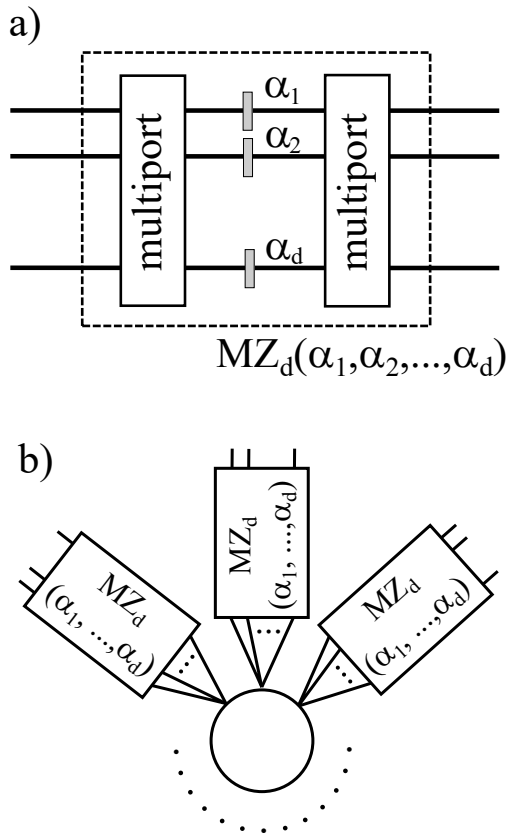}
    \caption{a). A generalised $d$-mode Mach-Zehnder (MZ) interferometer consisting of symmetric multiports intertwinned with $d$ phaseshifts to be estimated. b). An $N$-party configuration of the estimation setup, which consists of a central source of GHZ-path-entangled photons and $N$ local stations consisting of generalised MZ interferometers from Figure 1a). }
    \label{fig:setupl}
\end{figure}

\section*{Results}


\label{s1}

\subsection*{Analytical description of generalised Mach-Zehnder interferometer}

In this section, we describe our setup for estimation of multiple phases. For the convenience of the presentation, we will use the standard Hilbert space description, which demands the introduction of $N$ interferometers, however, as we discuss in details in Section \ref{sec:conc}, the entire scenario can be as well described using the second-quantised framework, in which only one interferometer suffices. The setup (see Figure 1b) consists of an $N$-partite GHZ source, which sends single photons to $N$ measurement stations, such that a $d$-mode bundle goes to each of the stations, and the basis states are encoded by a path of the photon within the bundle. Therefore the initial state reads:
\begin{eqnarray}
  &&\ket{\Psi^d}_N =U_d^{\otimes N}\ket{\textrm{GHZ}^d}_N=\nonumber\\
  &&\frac{1}{\sqrt{d}}  U_d^{\otimes N}(|\underbrace{0...0}_{N}\rangle +  |\underbrace{1...1}_{N}\rangle +   \ldots + |\underbrace{d-1...d-1}_{N}\rangle),
  \label{inState}
\end{eqnarray}
in which the preparation symmetric multiport $U_d$ will be specified later.
Further each of the $N$ measurement stations apply $d$-mode interferometers, which consist of generalised Mach-Zehnder interferometers \marcin{ involving $d$ phases, the $(d-1)$-element subsets of which will be estimated}. A generalised Mach-Zehnder interferometer \cite{Brougham2011} consists of two symmetric multiports intertwinned with a series of phaseshifts on each of the $d$ modes linking the multiports (see Figure 1a). 
\marcin{By a symmetric multiport we mean a $d$-mode multiport with the property that a single photon entering by any of the $d$ input modes has a uniform probability $\frac{1}{d}$ to be detected in any of the $d$ output modes (see e.g. \cite{Brougham2011}).}

For metrological considerations one usually needs a description of an interferometer in the Hamiltonian-like form $U=e^{-i h_i\alpha_i}$, with an explicit form of generators corresponding to phaseshifts. The typical choice of the operator basis for finding the generators $h_i$ is the set of Gell-Mann matrices, however we found that much more convenient choice for describing interferometers based on symmetric multiports is the Heisenberg-Weyl operator basis, defined as:
\begin{eqnarray}
\label{HWdef}
X &=& \sum_{i=0}^{d-1}|i\rangle \langle i+1 |,\nonumber\\
Z &=& \sum_{i=0}^{d-1} \omega^i|i\rangle \langle i|,\nonumber\\
Y &=&  XZ,
\end{eqnarray}
where $\omega = \exp(2i\pi/d)$. 
In the following we will present explicit form of the generators for $d=2,3,4$.
\subsubsection*{Two-mode case}
\marcin{In the two-mode case the Heisenberg-Weyl operators \eqref{HWdef} are equivalent to the standard Hermitian Pauli matrix basis.}
The symmetric two-port can be expressed in the following form:
\begin{equation}
\label{S2WH}
S_2 = \exp\left[ \frac{1}{4} i\pi X \right],
\end{equation}
whereas the phase-imprinting part of the interferometer has the following representation:
\begin{equation}
\label{F2WH}
F_2(\vec\alpha) = \exp\left[ \frac{1}{2}( \openone +  Z)i \alpha_{1} + \frac{1}{2}( \openone -  Z)i \alpha_{2}  \right].
\end{equation}
The evolution of the entire interferometer can be expressed in a concise way using only $Y$ operators:
\begin{eqnarray}
\label{U2}
\U_{2} &=& S_2F_2(\vec\alpha)(S_2)^{\dagger}= \exp\left[ \frac{1}{2}( \openone +  Y)i \alpha_{1} + \frac{1}{2}( \openone -  Y)i \alpha_{2}  \right].\nonumber\\
\end{eqnarray}
The $\U_2$ evolution is generated by two Hamiltonians:
\begin{eqnarray}
(\alpha_1 ~{\rm angle}) ~~ h_1&=&\frac{1}{2}(\openone +  Y),\\
(\alpha_2 ~{\rm angle}) ~~h_2&=&\frac{1}{2}(\openone - Y),
\end{eqnarray}
the eigenvalues of which read: $\{0,1\}$.\\

\subsubsection*{Three-mode case}
The description of the 3-mode case in the Heisenberg-Weyl basis turns out to be completely analogous to the 2-mode one. Firstly, the symmetric multiport has the following presentation in terms of generalised $X$ operators:
\begin{equation}
\label{S3WH}
S_3 = \exp\left[\frac{2}{9}i \pi ( X + X^2)\right],
\end{equation}
whereas the phase part depends solely on generalised $Z$ operators:
\begin{eqnarray}
\label{F3WH}
F_3(\vec\alpha) = \exp \bigg{[}\frac{1}{3}i \pi \left(\alpha_1 (Z + Z^2) + \alpha_2 ( \omega^2 Z + \omega Z^2) +
\alpha_3 ( \omega Z + \omega^2 Z^2)\right)\bigg].
\end{eqnarray}
The entire evolution is, in full analogy to the two-mode case, generated solely by the generalised $Y$ matrices:
\begin{eqnarray}
\label{U3}
\U_{3} = S_3F_3(\vec\alpha)(S_3)^{\dagger}=\exp\bigg[\frac{1}{3} i \left(\alpha_1 (\omega Y + \omega^2 Y^2) + \alpha_2 (Y + Y^2) +
\alpha_3 (\omega^2 Y + \omega Y^2) \right)\bigg].
\label{u3}
\end{eqnarray}
The $\U_3$ evolution is generated by three Hamiltonians:
\begin{eqnarray}
\label{h3}
(\alpha_1 ~{\rm angle}) ~~ h_1&=&\frac{1}{3}(\omega Y + \omega^2 Y^2),\nonumber\\
(\alpha_2 ~{\rm angle}) ~~h_2&=&\frac{1}{3}(Y + Y^2),\nonumber\\
(\alpha_3 ~{\rm angle}) ~~h_3&=&\frac{1}{3}(\omega^2 Y + \omega Y^2).
\end{eqnarray}
The above Hamiltonians fulfill $h_1+h_2+h_3 = 0$, and their eigenvalues read: $\{2/3,-1/3,-1/3\}$.
\marcin{Note that despite the fact that in the three-mode case the Heisenberg-Weyl operators \eqref{HWdef} are no longer Hermitian, their appropriate combinations give rise to a proper Hermitian Hamiltonians \eqref{h3}.}

\begin{figure}
\centering
	\includegraphics[width=0.48\textwidth]{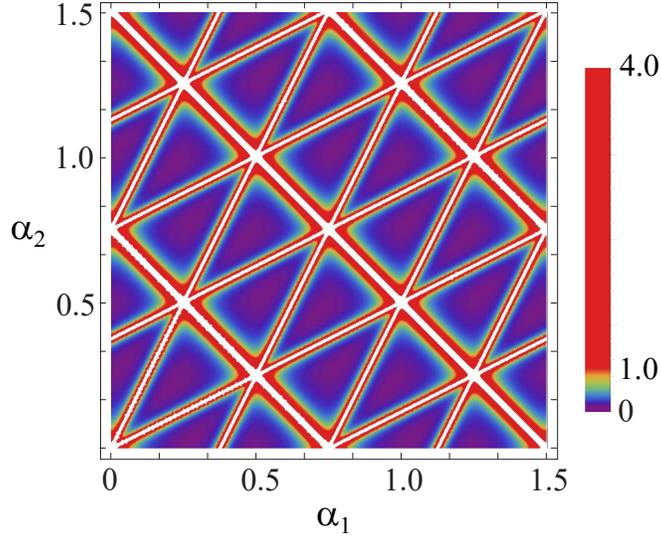}
	\caption{\label{Fig2}
\marcin{Plot of the right-hand-side of the Cramer-Rao bound \eqref{RCNN}, $\operatorname{Tr}\left(\left(\mathcal F(\alpha_1, \alpha_2)_{\mathcal I, \mathcal I}\right)^{-1}\right)$, as a function of the jointly estimated phases $\alpha_1$ and $\alpha_2$ in a $3$-mode Mach-Zehnder interferometer for the number of photons in the initial state set to $N=8$. The minimal value of $\operatorname{Tr}\left(\left(\mathcal F(\alpha_1, \alpha_2)_{\mathcal I, \mathcal I}\right)^{-1}\right)$ for the optimal values of the estimated phases reads $\frac{6+\sqrt 3}{128}\approx 0.06$, whereas the Monte-Carlo-estimated median, where both the phases were drawn uniformly, reads $0.65$.}}
\end{figure}

\subsubsection*{Four-mode case}
The description of a 4-mode case is a bit more complicated. The symmetric multiport still has analogous simple form in terms of generalised $X$ operator:
\begin{equation}
\label{S4WH}
S_4 =  \exp\left[\frac{1}{4} i \pi (X + X^2 + X^3)\right],
\end{equation}
whereas the phase-imprinting part can be presented in the following concise way:
\begin{eqnarray}
\label{F4WH}
F_4(\vec\alpha) = \exp[i \vec\alpha\cdot\vec z ],
\end{eqnarray}
where entries of the vector $\vec z$ fulfill relations:
\begin{equation}
\left(\begin{array}{c}z_1\\z_2\\z_3\\z_4\end{array}\right) = K \left(\begin{array}{c}\openone\\Z\\Z^2\\Z^3\end{array}\right),
\label{z}
\end{equation}
and $K_{mn} = \omega^{(m-1)(n-1)}$.
The entire evolution is generated by 4 Hamiltonians:
\begin{equation}
\mathcal{U}_{4} = S_4F_4(\vec\alpha)(S_4)^{\dagger}=\text{exp}\left[i (h_{1} + h_{2} + h_{3} + h_{4})  \right],
\label{u4}
\end{equation}
however their exact form, although still depending only on generalised $Y$ matrices,  is much more complicated than in previous cases:
\begin{eqnarray}
h_1& =& \frac{1}{4} \omega \left(Y - Y^2 + Y^3 -\eta ~{\rm Re} Y -
\theta ~ {\rm Re} Y^3\right),  \nonumber\\
h_2 &=& \frac{1}{4} \omega \left(-Y + Y^2 - Y^3 +\eta ~{\rm Re} Y + \eta ~ {\rm Re} Y^3\right),  \nonumber\\
h_3 &=& \frac{1}{4} \omega \left(Y^{\dagger} + (Y^3)^{*} - Y^2  - \mu {\rm Re} Y + \nu {\rm Re} Y^3\right), \nonumber \\
h_4 &=& \frac{1}{4} \omega \left(Y + Y^3 - (Y^2)^{*}  - \mu {\rm Re} Y + \nu {\rm Re} Y^3\right),
\label{h4}
\end{eqnarray}
where the star $^*$ denotes complex conjugation and  for clarity we introduced constants:
\begin{eqnarray}
\eta&=&\sqrt{\frac{2}{\omega}},\nonumber\\
\theta&=&\sqrt{2 \omega},\nonumber\\
\mu&=&2 \omega\left( 1 - \frac{i}{\sqrt{2 \omega}}\right),\nonumber\\
\nu&=&2 \omega\left( 1 - \frac{1}{\sqrt{2 \omega}}\right).\nonumber\\
\end{eqnarray}
The eigenvalues of $h_i$ are: $\{3/4, -1/4, -1/4, -1/4\}$. \marcin{In this case the same remark applies as in the previous one: properly defined functions of the Heisenberg-Weyl non-Hermitian operators give rise to  Hermitian Hamiltonians. }

\label{s3}

\subsection*{Precision of estimation of multiple phases with generalised Mach-Zehnder interferometer}

In this section, we analyse the estimation precision of our proposed estimation scheme for $d=3,4$ and an arbitrary number $N$ of photons in the initial state. There are two kinds of phases in our experiment: the $(d-1)$-element subset of unknown phases to be jointly estimated, and one remaining phase, assumed to be fixed and known, serving as a phase reference. \marcinNew{Such a situation, in which the set of all parameters determining the final probability distribution is divided into parameters of interest (the ones we estimate) and additional parameters, is well known in estimation theory \cite{Suzuki20} (see Methods section for more detailed presentation). In our case the additional parameter is the reference phase.} Since we assume entirely fixed experimental setup with no optimisation of measurements, we estimate the precision of estimation using classical Fisher Information Matrix techniques. \marcinNew{The precision of joint estimation of several parameters in the presence of fixed and known additional parameters is specified by the Cramer-Rao bound based on the inverese of the Fisher Information Submatrix corresponding to the parameters of interest \eqref{QRCNN}}. In order to describe the precision of estimation by a single quantity  we take the trace of both sides of the Cramer-Rao bound \eqref{QRCNN} \cite{Polino19}:
\begin{equation}
    \label{RCNN}
    \operatorname{Tr}(\operatorname{Cov}(\{A_{\mathcal I}\}))\geq \frac{\operatorname{Tr}\left(\left(\mathcal F(\vec \alpha)_{\mathcal I, \mathcal I}\right)^{-1}\right)}{\nu},
\end{equation}
where in our case the matrix $\mathcal F(\vec \alpha)_{\mathcal I, \mathcal I}$ is a Fisher Information submatrix corresponding to the subset of jointly estimated phases (denoted here by a symbol $\mathcal I$ meaning parameters of interest, see Methods). Therefore our task would be to analyse the behaviour of the quantity $\operatorname{Tr}\left(\left(\mathcal F(\vec \alpha)_{\mathcal I, \mathcal I}\right)^{-1}\right)$ as a function of the number of photons $N$ in order to find asymptotic scaling of precision.

In our setup we allow for an optimisation of the initial state, which depends solely on the dimension of the local multiport. As an optimisation strategy we take the maximisation of the mean QFI per parameter. We utilise the following inequality:
\begin{equation}
    \label{QFIineq}
    \frac{1}{d}\operatorname{Tr}\FQ\leq \frac{4}{d}\sum_{i=1}^d \langle H_i^2\rangle,
\end{equation}
which follows directly from \eqref{QFIMp} (see Methods section). Note that the LHS of the above inequality is just the mean QFI per parameter, which can be treated as an approximate measure of average estimation performance per parameter.
The total collective Hamiltonian corresponding to the action of $N$ local interferometers reads:
\begin{equation}
\label{ColHam}
H_i = h_i \otimes \openone \otimes ... \otimes \openone 
     + \openone \otimes h_i  \otimes \openone \otimes ... \otimes \openone  
     +  ... + \openone \otimes ... \otimes \openone \otimes h_i,
\end{equation}
where $h_i$ denotes any of the local Hamiltonians from formulas \eqref{h3} and \eqref{h4}.
The inequality \eqref{QFIineq} implies that the optimal state should be an eigenstate of the operator $\sum_i H_i^2$.

\subsubsection*{Three-mode case}
{\bf \em Optimal state.---}
The $N$-qutrit optimal state that maximizes the trace of the QFIM $\mathcal F^{Q}$ is given by:
\begin{equation}
|\Psi_3\rangle_N = \frac{1}{\sqrt{3}} U_{3}^{\otimes N}  (|0...0\rangle +  |1...1\rangle +   |2...2\rangle)
\label{optstate3},
\end{equation}
where  
\begin{equation}
 U_{3}=\frac{1}{\sqrt{3}} \left(
\begin{array}{ccc}
1 & 1 & 1 \\
\omega^2 & 1 & \omega \\
1 & \omega^2 & \omega \\
\end{array}
\right).   
\label{optu3}
\end{equation}
One can easily prove that fact by noticing that the operator $U_3$ simultaneously diagonalises the local Hamiltonians \eqref{h3}. Indeed, the operators \eqref{h3} are expressible solely by the operators $Y$ and $Y^2$, and we have the relations: $U_{3}^{\dagger} Y U_{3} = Z$ and $U_{3}^{\dagger} Y^2 U_{3} = Z^2$, where by definition $Z$ and $Z^2$ are diagonal. Consequently, following the action of the collective unitary operation $U_{3}^{\otimes N}$, the total collective Hamiltonians $H_i$ \eqref{ColHam} are  diagonal with the eigenstates $|0...0\rangle, |1...1\rangle$ and $|2...2\rangle$, which implies that the operator $\sum_i H_i^2$ is also diagonal with the same set of eigenstates.

{\bf \em Achievable Precision.---}
\marcin{As described in Methods section, in order to calculate the estimation precision via classical Fisher Information matrix \eqref{FIM} we have to determine the parameter-dependent probability distribution for measurement outcomes $p(k|\vec{\alpha})$. In our setup the outcomes are labeled by the numbers $i_k\in \{0,1,2\}$ which denote detector clicks in local modes $i_k$ in measurement stations $k$, therefore the distribution has the form $p(i_1,\ldots,i_N|\vec\alpha)$. Further, as we show in Methods section, the final probability distribution \eqref{finp3} depends only on the total number of clicks in local modes $\{0,1,2\}$ specified by integers $z,j,d$ respectively, therefore $p(i_1,\ldots,i_N|\vec\alpha)=p(z,j,d|\vec\alpha)$. Using the final form of the probability distribution \eqref{finp3}  we can directly calculate the classical $2\times 2$ Fisher Information submatrix \eqref{FIM} corresponding to joint estimation of two of the three phases $\{\alpha_1,\alpha_2\}=\mathcal I$, whereas the third phase is set to zero as a reference mode:
\begin{equation}
[\mathcal{F}(\alpha_1,\alpha_2)_{\mathcal I, \mathcal I}]_{ij} = \sum_{z,j,d=0}^N {N \choose z,j,d } \frac{\partial_{\alpha_i} p(z,j,d|\alpha_1,\alpha_2,0)\partial_{\alpha_j} p(z,j,d|\alpha_1,\alpha_2,0)}{p(z,j,d|\alpha_1,\alpha_2,0)},
\end{equation}
where the multinomial coefficient counts the number of separate detection situations giving rise to the total of $z,j,d$ clicks in modes $\{0,1,2\}$. In the above formula we took the third mode as a reference mode, however the form of the above Fisher information submatrix does not depend on this choice due to symmetry of the final probability distribution \eqref{finp3} with respect to the parameters $\alpha_i$. Therefore the following analysis holds for estimation of any 2-mode subset of modes of the interferometer.

The exact analytical expression for the above defined Fisher information submatrix for arbitrary values of $\alpha$'s is complicated, however we were able to calculate its inverse and find the optimal scaling of the quantity $\operatorname{Tr}\left(\left(\mathcal F(\vec \alpha)_{\mathcal I, \mathcal I}\right)^{-1}\right)$ as a function of the number of photons $N$. It reads:
\begin{equation}
    \operatorname{Tr}\left(\left(\mathcal F( \alpha_1^{\textrm{opt}},\alpha_2^{\textrm{opt}})_{\mathcal I, \mathcal I}\right)^{-1}\right)=\frac{6+\sqrt 3}{2N^2},
\end{equation}
for the optimal values of estimated phases:
\begin{equation}
\alpha_1^{\textrm{opt}}=\alpha_2^{\textrm{opt}}=\pi-\arctan\left(\sqrt{6+4\sqrt 3}\right)\approx 35.2^{\circ}.
\end{equation}
Assuming that the estimated phases are equal, $\alpha_1=\alpha_2=\alpha$, the trace of the inverse Fisher information submatrix has the following form:
\begin{equation}
    \operatorname{Tr}\left(\left(\mathcal F( \alpha)_{\mathcal I, \mathcal I}\right)^{-1}\right)=\frac{1}{N^2}\left(3+\cos(N\alpha)-\cos(2N\alpha)\right)(\csc(N\alpha))^2.
\end{equation}

In order to visualise how robust our strategy is for estimation of arbitrary values of the phases we plot the value of $\operatorname{Tr}\left(\left(\mathcal F( \alpha_1, \alpha_2)_{\mathcal I, \mathcal I}\right)^{-1}\right)$ as a function of the phases $\alpha_1,\alpha_2$ for $N=8$, see Figure \ref{Fig2}. 

\marcinNew{All the above analysis indicates the local character of the Fisher Information-based approach to precision of estimation: the precision strongly depends on the values of the estimated phases. Therefore in realistic applications one needs to obtain some prior knowledge of the phases in order to tune the interferometer in a way that the unknown phases are close to the optimal values for which the error is the lowest.}

Notice that even though we do not allow for optimization of final measurements, we still obtain the Heisenberg-like scaling of precision of joint estimation for each of the parameters around its optimal values.}

\subsubsection*{Four-mode case}
{\bf \em Optimal state.---}
In analogy to the previous case the $N$-ququart state which maximises the trace of the QFIM reads:
\begin{equation}
  \ket{\Psi_4}_N = \frac{1}{2} U_{4}^{\otimes N}  (|0...0\rangle +  |1...1\rangle +   |2...2\rangle + |3...3\rangle),
  \label{optstate4}
\end{equation}
where:
\begin{equation}
\label{optu4}
U_{4} = \frac{1}{2}
\left(
\begin{array}{cccc}
-1 & -1 & 1 & 1 \\
 1 & -1 & 1 & -1 \\
 1 & -1 & -1 & 1 \\
 1 & 1 & 1 & 1 \\
\end{array}
\right).
\end{equation}
The proof of optimality of \eqref{optstate4} follows the same logic as \eqref{optstate3}. $U_{4}$ simultaneosly diagonalizes local Hamiltonians $h_i$ \eqref{h4} with the eigenstates being the standard basis $|0\rangle, |1\rangle, |2\rangle$, and $|3\rangle $. Consequently, following the action of the unitary operation $U_{4}^{\otimes N} $, the total collective Hamiltonians $H_i$ \eqref{ColHam} are  diagonal with the eigenstates $|0...0\rangle, |1...1\rangle, |2...2\rangle$, and $|3...3\rangle$, which implies that the operator $\sum_i H_i^2$ is also diagonal with the same set of eigenstates.

{\bf \em Achievable Precision.---}
\marcin{In analogy to the previous case we have to determine the final probability distribution $p(k|\vec{\alpha})$. In the current case the outcomes are labeled by the numbers $i_k\in \{0,1,2,3\}$ which denote detector clicks in local modes $i_k$ in measurement stations $k$, therefore the distribution has the form $p(i_1,\ldots,i_N|\vec\alpha)$. As shown  in Methods section, the final probability distribution \eqref{finp4} depends only on the total number of clicks in local modes $\{0,1,2,3\}$ specified by integers $z,j,d,t$ respectively, therefore $p(i_1,\ldots,i_N|\vec\alpha)=p(z,j,d,t|\vec\alpha)$. Using the final form of the probability distribution \eqref{finp4}  we can directly calculate the classical $3\times 3$ Fisher Information submatrix corresponding to joint estimation of the three phases $\{\alpha_1,\alpha_2,\alpha_3\}=\mathcal{I}$, whereas the fourth phase is set to zero as a reference mode:
\begin{equation}
[\mathcal{F}(\alpha_1,\alpha_2,\alpha_3)_{\mathcal I, \mathcal I}]_{ij} = \sum_{z,j,d,t=0}^N {N \choose z,j,d,t } \frac{\partial_{\alpha_i} p(z,j,d,t|\alpha_1,\alpha_2,\alpha_3,0)\partial_{\alpha_j} p(z,j,d,t|\alpha_1,\alpha_2,\alpha_3,0)}{p(z,j,d,t|\alpha_1,\alpha_2,\alpha_3,0)},
\end{equation}
where the multinomial coefficient counts the number of separate detection situations giving rise to the total of $z,j,d,t$ clicks in modes $\{0,1,2,3\}$. As in the previous case the above Fisher information submatrix has the same form for any choice of the reference mode, therefore the following analysis of precision of estimation holds for estimating any triple of phases chosen from all the four ones.

Analogously to the previous case we calculated the optimal scaling of the quantity $\operatorname{Tr}\left(\left(\mathcal F(\vec \alpha)_{\mathcal I, \mathcal I}\right)^{-1}\right)$ as a function of the number of photons $N$. It reads:
\begin{equation}
    \operatorname{Tr}\left(\left(\mathcal F(\alpha_1^{\textrm{opt}}, \alpha_2^{\textrm{opt}},\alpha_3^{\textrm{opt}})_{\mathcal I, \mathcal I}\right)^{-1}\right)=\frac{6}{N^2},
\end{equation}
for the optimal values of estimated phases:
\begin{equation}
\label{optPhases4}
\alpha_1^{\textrm{opt}}=\alpha_2^{\textrm{opt}}=\alpha_3^{\textrm{opt}}=0.
\end{equation}
Assuming that all the estimated phases are equal, $\alpha_1=\alpha_2=\alpha_3=\alpha$, the trace of the inverse Fisher information submatrix scales with the number of photons $N$ as follows:
\begin{equation}
    \operatorname{Tr}\left(\left(\mathcal F( \alpha)_{\mathcal I, \mathcal I}\right)^{-1}\right)=\frac{3}{2N^2}\left(3+\cos(N\alpha)\right)\left(\sec\left(\frac{N\alpha}{2}\right)\right)^2.
\end{equation}

In this case, we also obtain the Heisenberg-like scaling of precision of joint estimation of any triple of the phases around their optimal values. \marcinNew{The same remark on the local character of precision of estimation applies here: one needs to gain a prior knowledge of the unknown phases in order to estimate them around the optimal working point specified by the optimal phases \eqref{optPhases4}}.
}
\label{s4}

\subsection*{Symmetric 5-mode multiport with symmetric Hamiltonian}

We found that symmetric multiports for $d=5,6$ can be also generated by the powers of generalised $X$ operators in analogy to formulas \eqref{S2WH}, \eqref{S3WH}, \eqref{S4WH}:
\begin{eqnarray}
\label{S5WH}
S_5&=&\exp\left[\frac{4\sqrt{5}}{25} i \pi (X - X^2 - X^3 + X^4)\right],\\
\label{S6WH}
S_6&=&\exp\left[i \pi \left(\frac{1}{3}X +\frac{1}{9} X^2 +\frac{1}{12} X^3 + \frac{1}{9}X^4+\frac{1}{3}X^5\right)\right].\nonumber\\
\end{eqnarray}
As pointed out in seminal paper \cite{Jex95} it is sometimes advisable to analyse the optical multiports from the Hamiltonian perspective. Such an analysis can be necessary in implementations of optical multiports with active optical devices. Let us move to the second quantisation description, in which we define the symmetric Hamiltonian for $d$-mode optical instrument as:
\begin{equation}
    H_{sym}=\sum_{i<j}e^{i\varphi_{ij}}(a_i^{\dagger}a_j+a_j^{\dagger}a_i),
    \label{symHam}
\end{equation}
which is a slight generalisation of the definition used in \cite{Jex95} which additionally includes phases.
In \cite{Jex95} (section III) it is stated that Hamiltonian of the form \eqref{symHam} cannot generate evolution of a symmetric multiport for $d>4$. However, it can be easily seen that the evolution  \eqref{S5WH} of a symmetric $5$-port is in fact generated by such a Hamiltonian. To see this let us notice that the Hamiltonian of the symmetric multiport \eqref{S5WH} reads up to a constant factor:
\begin{equation}
    H_5=X - X^2 - X^3 + X^4.
    \label{H5SW}
\end{equation}
Using the Jordan–Schwinger map:
\begin{equation}
    M\mapsto \sum_{i,j=1}^d a_i^{\dagger}M_{ij} a_j,
\end{equation}
which maps matrix operators on $\mathbb C^d$ into second-quantised operators on $d$-mode Fock space, we obtain that $H_5$ has the symmetric representation \eqref{symHam} with the following phases:
\begin{eqnarray}
\varphi_{12}&=&\varphi_{15}=\varphi_{23}=\varphi_{34}=\varphi_{45}=2\pi\nonumber\\
\varphi_{13}&=&\varphi_{14}=\varphi_{24}=\varphi_{25}=\varphi_{35}=\pi.
\end{eqnarray}
On the other hand the Hamiltonian:
\begin{equation}
    H_6=\frac{1}{3}X +\frac{1}{9} X^2 +\frac{1}{12} X^3 + \frac{1}{9}X^4+\frac{1}{3}X^5
    \label{H6SW}
\end{equation}
generating the evolution of the $6$-mode symmetric multiport \eqref{S6WH} does not have symmetric representation \eqref{symHam}. Instead it can be represented as:
\begin{equation}
    H_{6}=\sum_{i<j}\alpha_{ij}(a_i^{\dagger}a_j+a_j^{\dagger}a_i),
    \label{H6JS}
\end{equation}
with the following amplitudes:
\begin{eqnarray}
\alpha_{12}&=&\alpha_{16}=\alpha_{23}=\alpha_{34}=\alpha_{45}=\alpha_{56}=\frac{1}{3}\nonumber\\
\alpha_{13}&=&\alpha_{15}=\alpha_{24}=\alpha_{26}=\alpha_{35}=\alpha_{46}=\frac{1}{9},\nonumber\\
\alpha_{14}&=&\alpha_{25}=\alpha_{36}=\frac{1}{12}.
\end{eqnarray}

\subsection*{Single-interferometer optical implementation}
\label{sec:conc}
Although our scheme in the version presented in the Figure 1 can be directly implemented using optical interferometry, the main difficulty of such an implementation lies in preparing the multiphoton GHZ state. Despite the progress in realising multiphoton entanglement  in recent years (cf. eg. \cite{Hu2020}) it is still chalenging to prepare such states for higher number of subsystems, which may suggest that the proposed scheme is unfeasible. However our scheme can be transformed into simpler one which is completely feasible within current optical technology by performing second quantisation of the scheme. Note that the final states in our setup are symmetric states of photons which are distinguishable by path degree of freedom:
\begin{eqnarray}
\label{finStates}
\ket{\Psi_{out}^3}_N&=&\mathcal U_3^{\otimes N}U_3^{\otimes N}\ket{\rm{GHZ}^3}_N\nonumber\\
\ket{\Psi_{out}^4}_N&=&\mathcal U_4^{\otimes N}U_4^{\otimes N}\ket{\rm{GHZ}^4}_N.
\end{eqnarray}
The second-quantised version of the above states, which assumes that a state of  $N$ indistinguishable photons is send to a  single $d$-mode interferometer can be expressed as: 
\begin{eqnarray}
\label{finStates1}
\ket{\Psi_{out}^3}&=&\mathcal U_3U_3\ket{\rm{NOON}^3}_N\nonumber\\
\ket{\Psi_{out}^4}&=&\mathcal U_4U_4\ket{\rm{NOON}^4}_N,
\end{eqnarray}
where the $\ket{\textrm{NOON}^d}_N$ states are defined by the formula:
\begin{equation}
\label{noon}
    \ket{\textrm{NOON}^d}_N=\frac{1}{\sqrt{d}}  (|\underbrace{N0...0}_{d}\rangle +  |\underbrace{0N...0}_{d}\rangle +   \ldots + |\underbrace{0...0N}_{d}\rangle).
\end{equation}
The detection probabilities for the second-quantised version of the scheme are identical to the ones for the original scheme \eqref{finp3}, \eqref{finp4}, only the meaning of the detection events changes: now the numbers $z,j,d$ (and $t$ for the $4$-mode case) denote photon counts in modes $\{0,1,2\}$ (and respectively in mode $3$) in a \emph{single} interferometer consisting of the initial-state-correcting multiports $U_3$ \eqref{optu3} (or $U_4$ \eqref{optu4}) and the generalised Mach-Zehnder interferometer $\mathcal U_3$ \eqref{u3} (or $\mathcal U_4$ \eqref{u4}). Notice that the equivalence of the original scheme with the second-quantised one can be already seen at the level of probabilities \eqref{finp3} and \eqref{finp4}, since they do not distinguish in which of the $N$ stations there was a click in a given mode, but depend solely on the total number of clicks in given modes across all the labs. 

The single-interferometer version of the scheme is experimentally feasible at the current stage since, in contrast to the multiphoton GHZ sources, there are experimental methods to produce $N$-photon NOON states \eqref{noon}  for higher values of $N$ \cite{NooNGen}. \marcin{Note that a similar idea of using fixed multimode  Mach-Zehnder interferometer for multiphase estimation already appeared in several works \cite{Spagnolo12, Ciampini16, You17, Li20, Polino19}, however in all these works \marcinNew{input states with small definite number of photons} are considered, therefore they lack discussion about scaling of precision with the size of the initial state.}

\section*{Discussion}

In this work we have investigated the metrological properties of a generalised Mach-Zehnder interferometer for the number of modes equal to $3$ and $4$, with the emphasis put on the possibility of simultaneous estimation of $(d-1)$-element subset of phases placed in arbitrary configuration across the modes. We have shown that estimation of each of the subsets can be performed with Heisenberg-like scaling of precision in an \emph{entirely fixed interferometric setup}, namely with the same initial state and measurement strategy. To prove the Heisenberg-like scaling of precision we developed an analytical description of the generalised Mach-Zehnder interferometer in terms of Heisenberg-Weyl operators. This approach allows for analytical calculation of the inverse of the classical Fisher information matrix related with each of the subsets of parameters, which, in contrast to methods implicitly involving optimisation over measurement strategies based on Quantum Cramer-Rao or Holevo bounds, provides a \emph{factual} limit for the efficiency of estimation within assumed concrete measurement setup.


Since our scheme allows for estimation of any $(d-1)$-element subset of unknown phases placed arbitrarily across a fixed interferometer (the remaining phase is assumed to be known), the single-interferometer version of the scheme can be well suited for enhancing the performance of 3-dimensional quantum sensing tasks similar to the ones presented in \cite{Spagnolo12}, in which the estimation is performed using only \marcinNew{input states with a small definite number of photons}.

\section*{Methods}

\subsection*{General introduction to multiphase estimation}

Standard approach to multiparameter estimation assumes the following estimation scheme \cite{SzczykulskaRev, Gessner18, Ragy16}: an initial multipartite state $\rho_{in}$ undergoes an evolution $\Lambda_{\vec\alpha}$, which depends on a vector of unknown parameters $\vec \alpha=(\alpha_1,\ldots,\alpha_d)$. Finally single-particle projective measurements $\{\Pi\}_k$ with outcomes labeled as $k$'s are performed, leading to final probability distribution:
\begin{equation}
p(k|\vec\alpha)=\operatorname{Tr}(\Lambda_{\vec\alpha}(\rho_{in})\Pi_k).
\end{equation}
Having the parameter-dependent probability distribution $p(k|\vec\alpha)$ one can construct estimators $\{A_i\}$ of the unknown parameters $\{\alpha_i\}$. In order to estimate the joint accuracy of these estimators one has to introduce joint measure of sensitivity of the distribution $p(k|\vec\alpha)$ on the parameters  $\{\alpha_i\}$. In classical estimation theory such a measure is provided by the Fisher Information Matrix $\F$ (FIM), defined as:
\begin{equation}
\F_{ij}(\vec \alpha)=\sum_k \frac{\partial_{\alpha_i} p(k|\vec \alpha)\partial_{\alpha_j} p(k|\vec \alpha)}{p(\vec k|\vec \alpha)}.
\label{FIM}
\end{equation} 
If the estimators  $\{A_i\}$ are unbiased, namely their mean values equal to $\{\alpha_i\}$ for the entire range of $\alpha$'s, and the $\F$ matrix is invertible, the quality of estimation of $\{\alpha_i\}$ based on distribution $p(k|\vec\alpha)$ is described by the Cramer-Rao bound:
\begin{equation}
    \label{CRC}
    \operatorname{Cov}(\{A_i\})\geq \frac{\F^{-1}}{\nu},
\end{equation}
where $\operatorname{Cov}(\{A_i\})$ is a covariance matrix for estimators, namely $\operatorname{Cov}(\{A_i\})_{mn}=\operatorname{Cov}(A_m,A_n)$, and $\nu$ is the number of repetitions of the experiment. The above description of the efficiency of 
estimation assumes fixed measurements $\{\Pi\}_k$.
In quantum estimation theory one is usually interested in description of efficiency of estimation of $\alpha$'s from an evolved quantum state $\rho_{out}(\vec\alpha)=\Lambda_{\vec\alpha}(\rho_{in})$ in a way which assumes optimisation over all possible measurements. This idea is encoded in the Quantum Fisher Information Matrix (QFIM), defined in an operator-based way:
\begin{equation}
    \label{QFIM}
    \FQ_{ij}=\frac{1}{2}\operatorname{Tr}(\rho_{out}(\vec\alpha)\{L_i,L_j\}),
\end{equation}
where the braces denote anticommutator of operators and the operators $L_i$ are defined implicitly by the equation:
\begin{equation}
    \label{SLD}
   \frac{1}{2}\{L_i,\rho_{out}(\vec\alpha)\}=\partial_{\alpha_i}\rho_{out}(\vec\alpha).
\end{equation}
In the case of pure input state $\ket{\psi}$ and the unitary evolution of the form $U=e^{-iH_i\alpha_i}$ the QFIM can be expressed in an explicit form:
\begin{equation}
    \label{QFIMp}
    \FQ_{ij} = 4 \langle {\rm cov} H \rangle_{\psi} = 4 \left(\langle H_i H_j \rangle_{\psi}-\langle H_i \rangle_{\psi}\langle H_j \rangle_{\psi}\right).
\end{equation}
Assuming that the QFIM is invertible the Quantum Cramer-Rao bound holds:
\begin{equation}
    \label{QRC}
    \operatorname{Cov}(\{A_i\})\geq \frac{(\FQ)^{-1}}{\nu}.
\end{equation}

\marcin{It is worth to mention that there exists another version of the Quantum Cramer-Rao bound, namely the Holevo bound \cite{Holevo82, Ragy16, Datta20}, which does not utilise the QFIM.} This bound on the precision of estimation is expressed using the notion of a cost matrix $\mathcal C$, which is a positive matrix providing weights to the uncertainties related with different parameters and the matrix $\mathcal V$ defined elementwise by the relation $\mathcal V_{ij}=\operatorname{Tr}(\mathcal X_i\mathcal X_j\rho_{out}(\vec\alpha))$, where the Hermitian matrices $\mathcal X_i$ fulfill the constraint $\operatorname{Tr}(\{\mathcal X_i,L_j\}\rho_{out}(\vec\alpha))=2\delta_{ij}$  and the operators $L_i$ are defined in \eqref{SLD}. Then the Holevo bound has the following form \cite{Ragy16}:
\begin{equation}
    \label{Holevo}
    \operatorname{Tr(\mathcal{C}\operatorname{Cov}(\{A_i\}))\geq \min_{\mathcal X_i}\left(\operatorname{Tr(\mathcal C \operatorname{Re}(\mathcal V))}+||\sqrt{\mathcal C}\operatorname{Im}(\mathcal V)\sqrt{\mathcal C}||\right)},
\end{equation}
where the norm in the RHS is the trace norm. The Holevo bound \eqref{Holevo} is tighter than the QCRB bound \eqref{QRC}. \marcin{Despite the fact that the Holevo bound does not utilise the QFIM, it is also ill-defined in the case when the corresponding QFIM-based Cramer-Rao bound is ill-defined due to singularity of the QFIM \cite{Datta20}.}

\marcinNew{
\subsection*{Multiparameter Cramer-Rao bound in the presence of additional parameters}}

Application of both the bounds \eqref{QRC} and \eqref{Holevo} always raises the question whether found precision limitations can be saturated by experimentally accessible measurement schemes. 
Our main aim is to investigate the process of estimation of multiple phases within a \emph{fixed} interferometer and a simple fixed measurement scheme consisting of single output mode measurements.
Therefore our approach to evaluate the precision of estimation would be based on basic tools in estimation theory, namely on the classical Fisher information matrix techniques. In this way we do not need to care about optimisation of measurements. 

However another issue remains to be solved concerning our setup. Namely our task is to estimate multiple phases, which constitute a subset of all the parameters on which the final probability distribution depends (being the phases to be estimated and the reference phase). \marcinNew{Such a situation in estimation theory has been discussed extensively in \cite{Gross20, Suzuki20}. In our case the chosen subset of phases is referred to as the set of parameters of interest, whereas the reference phase is known and fixed. In general two different cases have to be considered when additional parameters than the estimated ones appear in the setup: (i) the additional parameters are fixed and known, (ii) the additional parameters are fixed but unknown (and are called in this context the \emph{nuisance parameters}). Let us denote the division of the set of all parameters into parameters of interest and additional parameters by a vector $(\vec \alpha_{\mathcal I}, \vec \alpha_{\mathcal O})$ in the case additional parameters are known and as $(\vec \alpha_{\mathcal I}, \vec \alpha_{\mathcal N})$ if they are the nuisance parameters. Then the Fisher Information Matrix and its inverse can be expressed in a block form with respect to the fixed partition of parameters $(\vec \alpha_{\mathcal I}, \vec \alpha_{\mathcal O})$ and $(\vec \alpha_{\mathcal I}, \vec \alpha_{\mathcal N})$ \cite{Suzuki20}:
\begin{eqnarray}
 &&\mathcal{F}(\vec \alpha) = \left(
\begin{array}{cc}
 \mathcal F(\vec \alpha)_{\mathcal I, \mathcal I} &  \mathcal F(\vec \alpha)_{\mathcal I, \mathcal O} \\
  \mathcal F(\vec \alpha)_{\mathcal O, \mathcal I} &  \mathcal F(\vec \alpha)_{\mathcal O, \mathcal O} \\
\end{array}
\right) \\
&&\left(\mathcal{F}(\vec \alpha)\right)^{-1} = \left(
\begin{array}{cc}
 \mathcal G(\vec \alpha)_{\mathcal I, \mathcal I} &  \mathcal G(\vec \alpha)_{\mathcal I, \mathcal N} \\
  \mathcal G(\vec \alpha)_{\mathcal N, \mathcal I} &  \mathcal G(\vec \alpha)_{\mathcal N, \mathcal N} \\
\end{array}
\right).
\end{eqnarray}}

\marcinNew{The Cramer-Rao bound for the parameters of interest on condition that the additional parameters are known has the following form:
\begin{equation}
    \label{QRCNN}
    \operatorname{Cov}(\{A_{\mathcal I}\})\geq \frac{\left(\mathcal F(\vec \alpha)_{\mathcal I, \mathcal I}\right)^{-1}}{\nu},
\end{equation}
whereas for the case they are unknown (they are the nuisance parameters) it reads:
\begin{equation}
    \label{QRCNU}
    \operatorname{Cov}(\{A_{\mathcal I}\})\geq \frac{\mathcal G(\vec \alpha)_{\mathcal I, \mathcal I}}{\nu}.
\end{equation}
In simple words the covariance of the estimators of parameters of interest is bounded from below either by the \emph{inverse of the submatrix} of the FIM (when additional parameters are known) or by the \emph{submatrix of the inverese} of the FIM (if they are unknown). It is worth to mention, that all of this holds also in the case of a single parameter of interest: if the additional parameters are unknown, still one needs to take into account entire FIM and take its inverse \cite{Gross20}.}
\label{s2}

\subsection*{Probability distributions for the particular outcomes of the experimental setup}

\label{appenA}
\subsubsection*{Three-mode case}
\label{asub1}
Let $i_k\in\{0,1,2\}$ denote the measurement outcome at $k$-th station corresponding to detector click in $i_k$-th local mode. Then the conditional probability distribution for the outcomes conditioned on the values of the phaseshifts reads:
	\begin{eqnarray}
	\label{app3}
		p(i_1,i_2,...,i_N|\alpha_1,\alpha_2,\alpha_3)
		&=& |\langle i_1 \dots i_N| ~  \mathcal{U}_{3}(\alpha_1,\alpha_2,\alpha_3)^{\otimes N} U_{3}^{\otimes N} ~|GHZ\rangle_N         |^2\nonumber\\
		&=&\frac{1}{3} \bigg|\sum_{j = 0}^{2}\langle i_1 \dots i_N|   ~\mathcal{U}_{3}(\alpha_1,\alpha_2,\alpha_3)^{\otimes N} U_{3}^{\otimes N}~ |\underbrace{j\dots j}_{N}\rangle      \bigg|^2 \nonumber\\
				&=&\frac{1}{3} \bigg|\sum_{j = 0}^{2} {\langle i_1 \dots i_N|   ~\mathcal{U}_{3}(\alpha_1,\alpha_2,\alpha_3). U_{3} ~|j\rangle}^{N}  \bigg|^2 \nonumber\\
					&=&\frac{1}{3} \left|\sum_{j = 0}^{2} \prod_{k=1}^N \langle i_k| ~ \mathcal{U}_{3}(\alpha_1,\alpha_2,\alpha_3) .U_{3} ~|j\rangle]  \right|^2 
					\label{pn3}
	\end{eqnarray}
Firstly the entire local evolution operator $\mathcal{U}_{3}(\alpha_1,\alpha_2,\alpha_3). U_{3}$ can be presented in a compact matrix form by direct use of the defining formulas \eqref{u3} and \eqref{optu3}:
\begin{equation}
 \mathcal{U}_{3}(\alpha_{1},\alpha_{2},\alpha_{3}).U_{3} = \frac{1}{\sqrt{3} }\left(
\begin{array}{ccc}
 e^{i \alpha_{2}} & e^{i \alpha_{3}} & e^{i \alpha_{1}} \\
 \omega^{2} e^{i \alpha_{2}} & e^{i\alpha_{3}} & \omega e^{i \alpha_{1}} \\
 e^{i \alpha_{2}} & \omega^{2} e^{i \alpha_{3}} & \omega e^{i \alpha_{1}} \\
\end{array}
\right) e^{- \frac{1}{3} i (\alpha_{1}+\alpha_{2}+  \alpha_{3})}.
\end{equation}
Secondly it turns out that the probability distribution \eqref{app3} has an important symmetry, namely it depends only on the total number of local clicks in local modes $\{0,1,2\}$, which we denote by $z,j,d$ respectively. Using this property the final form of the probability distribution reads:
\begin{eqnarray}
\label{finp3}
p(\underbrace{0, \cdots,  0}_{z}, \underbrace{1, \cdots, 1}_{j},  \underbrace{2, \cdots, 2}_{d}| \alpha_{1},\alpha_{2},\alpha_{3}) &=& \frac{1}{3^{N+1}}  \Bigg(  3  + 2 \cos{\left[ \frac{2\pi}{3}(d-j) + N(\alpha_{1} - \alpha_{2})\right] } \nonumber \\ \Bigg.&+& ~ 2 \cos{\left[\frac{2\pi}{3}(j-d) + N(\alpha_{1} - \alpha_{3})\right] } \nonumber \\ ~ &+& ~ \Bigg. 2 \cos{\left[\frac{4\pi}{3}(j-d) + N(\alpha_{2} - \alpha_{3})\right] } 
\Bigg).
\end{eqnarray}
From here, it is very easy to derive compact forms of the elements of the Fisher Information Matrix. 

\subsubsection*{Four-mode case}
\label{asub2}
Again, using the same logic as in the three-mode case, the formula for conditional probability of detection events for $d = 4$ has the following form:
\begin{equation}
		p(i_1,i_2,...,i_N|\alpha_1,\alpha_2,\alpha_3,\alpha_4) = \frac{1}{4} \left|\sum_{j = 0}^{3} \prod_{k=1}^N \langle i_k|   \mathcal{U}_4(\alpha_1,\alpha_2,\alpha_3,\alpha_4).U_{4} |j\rangle]  \right|^2.
	\end{equation}
In full analogy to the previous case the local evolution operator $\mathcal{U}_4(\alpha_1,\alpha_2,\alpha_3,\alpha_4).U_{4}$ has a compact matrix form:
\begin{equation}
 \mathcal{U}_{4}(\alpha_{1},\alpha_{2},\alpha_{3},\alpha_{4}).U_{4} =  \frac{1}{2}\left(
\begin{array}{cccc}
 -e^{i \alpha_{1}} & -e^{i \alpha_{2}} & e^{i \alpha_{3}} & e^{i \alpha_{4}} \\
 e^{i \alpha_{1}} & -e^{i \alpha_{2}} & e^{i \alpha_{3}} & -e^{i \alpha_{4}} \\
 e^{i \alpha_{1}} & -e^{i \alpha_{2}} & -e^{i \alpha_{3}} & e^{i \alpha_{4}} \\
 e^{i \alpha_{1}} & e^{i \alpha_{2}} & e^{i \alpha_{3}} & e^{i \alpha_{4}} \\
\end{array}
\right) e^{-\frac{1}{4} i (\alpha_{1} + \alpha_{2} + \alpha_{3} + \alpha_{4})}.
\end{equation}
The final probability distribution again depends only on the total number of clicks in local modes $\{0,1,2,3\}$ denoted respectively as $z,j,d,t$:
\begin{eqnarray}
\label{finp4}
&&p(\underbrace{0, \cdots,  0}_{z}, \underbrace{1, \cdots, 1}_{j},  \underbrace{2, \cdots, 2}_{d},\underbrace{3, \cdots, 3}_{t}| \alpha_{1},\alpha_{2},\alpha_{3},\alpha_{4}) \nonumber\\&&= \frac{1}{4^{N+1}}  \Big(  4  + 2 \cos{\left[ (d + j) \pi + N(\alpha_{2} - \alpha_{1})\right] }  +  2 \cos{\left[(d - z) \pi + N(\alpha_{3} - \alpha_{1})\right] } \Big.\nonumber \\ && \Big.+ 2 \cos{\left[(j - z) \pi + N(\alpha_{4} - \alpha_{1})\right] }  + 2 \cos{\left[(j + z) \pi + N(\alpha_{2} - \alpha_{3})\right] } \nonumber \\ &&\Big. + 2 \cos{\left[(d + z) \pi + N(\alpha_{2} - \alpha_{4})\right] } +  2 \cos{\left[(d - j) \pi + N(\alpha_{3} - \alpha_{4})\right] } 
\Big).
\end{eqnarray}

\section*{Acknowledgements}

The authors acknowledge discussions with Lukas Knips and Aaron Goldberg.
The work is part of the ICTQT IRAP (MAB) project of the Foundation for Polish Science (IRAP project
ICTQT, Contract No. 2018/MAB/5, cofinanced by EU via Smart Growth Operational Programme). M.P. acknowledges the support by DFG (Germany) and NCN (Poland) within the joint funding initiative Beethoven2 (Grant No. 2016/23/G/ST2/04273).

\bibliography{genMZ}

\end{document}